\documentstyle[prl,aps,multicol,epsf]{revtex}

\begin{document}
\draft
\widetext

\title{Scattering approach to classical quasi-1D transport}

\author{Eugene Kogan}
\address{Jack and Pearl Resnick Institute 
of Advanced Technology,\\
Department of Physics, Bar-Ilan University, Ramat-Gan 52900, 
Israel}

\date{\today}
\maketitle

\widetext
\begin{abstract}
\leftskip 54.8pt
\rightskip 54.8pt

General dynamical  transport of classical particles
in disordered quasi-1D samples is viewed in the framework of
scattering approach. Simple equation for the transfer-matrix is
obtained within this unified picture. In the case of diffusive
transport  the solution of this equation exactly
coincides with the solution of diffusion equation.

\end{abstract}


\begin{multicols}{2}
\narrowtext

The problem of classical particles transport in a disordered
quasi-1D system has a long history. In fact, many would claim
that the problem itself is history. I want to show in this paper
that this is not the case, and that the well known problem can be
seen in a new light. 

We consider the  quasi-1D system. Classical particle is
impinging  from the left and  leaves the sample after the time
$t$  from the input face with the differential probability $R(t)$
and from the  output face with the differential probability
$T(t)$.  We want to calculate these probability functions. The
problem can be reduced to finding, for any given frequency, two
Fourier transforms -- $T_{\omega}$, which we shall call 
transmission coefficient,  and $R_{\omega}$, which we shall call
reflection coefficient (both of course are frequency dependent). 

We are going to show that  scattering approach which is
extensively used in quantum transport, can serve as simple and
giving physical insight tool to describe the classical problem
also. In the framework of this approach  the transport of
particles through the sample can be represented as scattering
from one of the two input channels  into either one of the two
output channels (see Fig.1).

\begin{figure}
\epsfxsize=2.5truein
\centerline{\epsffile{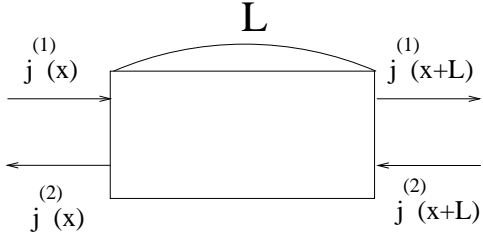}}
\caption{General setup for transport experiment}
\label{Fig.1}
\end{figure}

The same process we can  represent introducing the notion of
transfer matrix $M$, which connect  the differential
probabilities on the left side  $(j_1(x)$ and $j_2(x))$
with the differential
probabilities on the right side $(j_1(x+L)$ and $j_2(x+L))$
\begin{equation}
\label{matr}
\hat{j}_{\omega}(x+L) = \hat{M}_{\omega}(L) \cdot 
\hat{j}_{\omega}(x),
\end{equation}
where
\begin{equation}
\label{current}
\hat{j}=\left( 
\begin{array}{c}
j^{(1)}   \\ 
j^{(2)} 
\end{array}
\right).
\end{equation}
The  transfer-matrix give full description of the transport; 
transmission and reflection coefficients are simply connected 
with its matrix elements

\begin{equation}
\label{m}
\hat{M}_{\omega}=\left( 
\begin{array}{cc}
\frac{1}{T_{\omega}}& -\frac{R_{\omega}}{T_{\omega}}   \\ 
\frac{R_{\omega}}{T_{\omega}}& T_{\omega}-
\frac{R_{\omega}^2}{T_{\omega}} 
\end{array}
\right). 
\end{equation}

To obtain the equation for the transfer-matrix let us consider
$L$ not just as a fixed length of  the sample but as a free
parameter.  Now let us imagine $L$ being divided into two
arbitrary parts $L_1$ and $L_2$.

\begin{equation}
\label{sum}
L=L_1+L_2,
\end{equation} 
The convenience of using the
transfer-matrix  is due to the fact  that the transfer-matrix of
the whole sample is the
product of the transfer-matrices of its parts

\begin{equation}
\label{product}
\hat{M}_{\omega}(L_1+L_2)=\hat{M}_{\omega}(L_1)\cdot \hat{M}_{\omega}(L_2).
\end{equation} 
This equation is quite general. It is valid provided only  the
sample is homogeneous. (When the system is not in the multiple
scattering regime Eq.(\ref{product}) implies also 
the condition of the "diffusive illumination").   The
Eq.(\ref{product}) looks like a symbolic one. But we'll show
that it can be really solved, and e.g. for a diffusive transport
its solution exactly coincides with that of the diffusion
equation with the appropriate boundary conditions.

Let us first consider the case of zero frequency. In this case,
taking  into account the conservation law
\begin{equation}
\label{conservation}
R(L)+T(L)=1,
\end{equation}
we can rewrite Eq.(\ref{product})  in the form
\begin{equation}
\label{ohm}
\frac{R}{T}(L_1+L_2)= \frac{R}{T}(L_1)+ \frac{R}{T}(L_2).
\end{equation}
The solution is obvious
\begin{equation}
\label{statsol}
T(L)=\frac{1}{1+\alpha L},
\end{equation}
where $\ell^{\ast}$ is some constant, which is  determined
by the equation
\begin{equation}
\label{statdif}
R(dL)=1-T(dL)=\alpha dL.
\end{equation}

Now let us return to the general case of finite frequency.
In this case it is convenient to write Eq.(\ref{product}) 
in differential form
\begin{equation}
\label{matrix}
\frac{d \hat{M}_{\omega}(L)}{d L}= \hat{K}_{\omega}\cdot \hat{M}_{\omega}(L),
\end{equation}
where 
\begin{equation}
\label{k}
\hat{K}_{\omega}=\frac{d\hat{M}_{\omega}}{d L} {\Big|}_{L=0}.
\end{equation}
This equation should be supplemented by the boundary condition
\begin{equation}
\label{boundary3}
\hat{M}_{\omega}(0)=\left( 
\begin{array}{cc}
1& 0 \\ 
0&1 
\end{array}
\right),
\end{equation}
Eq.(\ref{matrix})
expresses transfer-matrix through two constants, defined by the
equations
\begin{equation}
\label{dyndif}
T_{\omega}(dL)=1-\alpha_{\omega} dL,\;\;
R_{\omega}(dL)=\beta_{\omega} dL.
\end{equation}
Thus
\begin{equation}
\label{k2}
\hat{K}_{\omega}=\left( 
\begin{array}{cc}
\alpha_{\omega}& -\beta_{\omega}   \\ 
\beta_{\omega}&-\alpha_{\omega} 
\end{array}
\right). 
\end{equation}

Inserting  $\hat{K}_{\omega}$ into Eq.(\ref{matrix}) and using the 
boundary conditions (\ref{boundary3}) we get
\begin{equation}
\label{sol}
\hat{M}_{\omega}(L)=\frac{\beta_{\omega}}{C_{\omega}}\left( 
\begin{array}{cc}
\sinh{\left[C_{\omega}(L+\ell_{\omega})\right]}&
 -\sinh {(C_{\omega} L)}   \\ 
\sinh {(C_{\omega} L)}&
-\sinh{\left[C_{\omega}(L-\ell_{\omega})\right]} 
\end{array}
\right), 
\end{equation}
where 
\begin{equation}
\label{c}
C_{\omega}= \sqrt{\alpha_{\omega}^2-\beta_{\omega}^2},
\end{equation}
and $\ell_{\omega}$ is found from the equation
\begin{equation}
\label{lomega}
\beta_{\omega} \sinh{(C_{\omega}\ell_{\omega})}= C_{\omega}
\end{equation}

Eq.(\ref{sol}) is the main result of the paper. We see that the
Fourier components of the transmission and reflection
coefficients have universal $L$ - dependence, whatever the
character of the transport and the detailed microscopics of the
system is. The $\omega$-dependence of the Fourier components is
less universal, so to say something meaningful about it we should
specify the character of the transport.

For illustration  consider the diffusive transport.
Let us start from the static case.
To calculate $\alpha$ we should consider the
transmission through the part of the
sample $dL$ much less than the
mean free path $\ell$ at the kinetic level of description, using,
say Boltzmann equation.
But even without further
reasoning it is obvious that 
\begin{equation}
\label{st2}
R(dL)=1-T(dL)=\frac{k dL}{\ell},
\end{equation}
where k is some numerical coefficient of the order of one, exact
value of which depends upon the microscopics of the system
(similarly to the fact that the exact form of boundary conditions
to the diffusion equation depends upon microscopics of the
system). Hence we see that   apart from numerical coefficient of
the order of one  $\alpha$ is just the inverse mean
free path. 

Let us consider now the case of finite frequency. 
In this case we can connect $\alpha_{\omega}$ and
$\beta_{\omega}$ (more exactly their difference) with the
diffusion coefficient $D$.  
Eq.(\ref{matr}) in the differential form can be written as
\begin{equation}
\label{matrd}
\frac{\partial \hat{j}_{\omega}}{\partial x} = \hat{K}_{\omega}(L)\cdot 
\hat{j}_{\omega}.
\end{equation}
For the physical current $j$ 
\begin{equation}
\label{phys}
j=j^{(1)}-j^{(2)}
\end{equation} 
we  obtain
\begin{equation}
\label{matrdd}
\frac{\partial^2 j_{\omega}}{\partial x^2} = 
\left(\alpha_{\omega}^2-\beta_{\omega}^2\right) j_{\omega}.
\end{equation}
Comparing this equation with the diffusion equation we
immediately obtain
\begin{equation}
\label{diffusion3}
C_{\omega}=\alpha_{\omega}^2-\beta_{\omega}^2 = \frac{i \omega}{D}.
\end{equation}
On the other hand in diffusive approximation
we restrict ourselves
by the leading order approximation with respect to
small parameter 
\begin{equation}
\label{un}
\sqrt{\frac{i \omega}{D}}\ell \ll 1.
\end{equation}
It means that  for the diffusive regime
we should put into Eq.(\ref{sol})
\begin{equation}
\label{diffusion2}
\beta_{\omega} = \ell_{\omega}^{-1}=\beta_0=\alpha.
\end{equation}
After inserting these constants  we obtain for
the transmission and reflection coefficient results which exactly
coincide with those  following from the diffusion equation
\cite{morse}.

\end{multicols}
\end{document}